\documentclass[eqsecnum,aps,twocolumn,epsf,showpacs]{revtex4} 
\usepackage{graphicx}
 
\begin{document}
   
 

\title{Fermionic bright  soliton  in
a  boson-fermion mixture}
 
\author{Sadhan K. Adhikari\footnote{Electronic
address: adhikari@ift.unesp.br; \\
URL: http://www.ift.unesp.br/users/adhikari/}}
\affiliation
{Instituto de F\'{\i}sica Te\'orica, Universidade Estadual
Paulista, 01.405-900 S\~ao Paulo, S\~ao Paulo, Brazil\\}
{Accepted:  Physical Review A} 
 
\date{\today}
 
 
\begin{abstract}

We use a time-dependent dynamical mean-field-hydrodynamic model to study
the formation of fermionic bright 
solitons in a trapped 
degenerate Fermi gas mixed with a Bose-Einstein
condensate in a quasi-one-dimensional cigar-shaped geometry. Due to a
strong
Pauli-blocking repulsion among 
spin-polarized fermions at short distances 
there cannot be bright fermionic
solitons in the case of repulsive boson-fermion interactions. 
However,  we demonstrate that stable bright fermionic
solitons can be formed for a sufficiently attractive boson-fermion
interaction in a boson-fermion mixture. We also consider the formation of
fermionic solitons in the
presence of a periodic  axial optical-lattice potential.
These solitons can be formed and studied in the laboratory
with present
technology.

\end{abstract}

\maketitle

\section{Introduction}

There cannot be an  effective evaporative
cooling  leading to a trapped quantum degenerate  Fermi gas (DFG) 
due to a strong repulsive Pauli-blocking interaction at low
temperature  among spin-polarized fermions \cite{exp1}. 
However, it has been possible to achieve a DFG by sympathetic cooling in the presence of a 
second
boson or fermion
component.  
Recently, there have been  successful observation
\cite{exp1,exp2,exp3,exp4} and associated  experimental
\cite{exp5,exp5x,exp6} and theoretical \cite{yyy1,yyy,zzz,capu,ska} 
studies of mixtures of a trapped DFG and a Bose-Einstein condensate
(BEC)   by different experimental groups
\cite{exp1,exp2,exp3,exp4} in 
the following systems: $^{6,7}$Li \cite{exp3}, $^{23}$Na-$^6$Li
\cite{exp4} and 
$^{87}$Rb-$^{40}$K \cite{exp5,exp5x}. 
Also,  there have been 
studies of mixtures of two-component trapped DFGs in 
$^{40}$K \cite{exp1} and  
$^6$Li \cite{exp2} atoms.

The formation and collapse 
of a DFG in a boson-fermion mixture
$^{87}$Rb-$^{40}$K 
have been
observed and studied by Modugno {\it et al.}
\cite{exp5,zzz,ska}. Although, the fermion-fermion interaction at short
distances 
is
repulsive due to strong Pauli blocking 
and hence incapable of leading to collapse,  
a sufficiently  
attractive boson-fermion interaction could overcome the Pauli repulsion
and could result in a collapse of a DGF.
Bright solitons in a BEC are formed due to an
attractive nonlinear atomic interaction. 
As the interaction in a pure
DFG at short distances is repulsive, 
there cannot be
bright solitons in 
a DFG. 

In this paper 
we study the possibility of the formation of stable fermionic bright
solitons  
in a mixture of a DFG with a BEC   in the presence of a sufficiently attractive
boson-fermion interaction which can overcome the Pauli repulsion among 
fermions. 
The formation of a fermionic  soliton is related to the fact 
that the system can lower its energy by forming high density regions
(the solitons) when the   attraction between the bosons and fermions is
large enough to overcome the Pauli repulsion in the DFG and any possible 
repulsion in the BEC. 
In particular we consider 
the formation of fermionic bright  solitons, which can freely move in the
axial direction,  in
such a mixture for  a quasi-one-dimensional cigar-shaped 
geometry
using a coupled time-dependent mean-field-hydrodynamic
model where the  bosonic component is treated by the mean-field
Gross-Pitaevskii 
equation \cite{11} and the fermionic component is treated by a
hydrodynamic
model \cite{capu}. 
This time-dependent mean-field-hydrodynamic model was
suggested
recently by the present author \cite{ska} to study the collapse 
dynamics
of fermions and is a
time-dependent extension of a time-independent model used for the
stationary  states by Capuzzi {\it et al.} \cite{capu}.

Bright solitons are really eigenfunction of the one-dimensional nonlinear
Schr\"odinger equation. However, the experimental realization of bright
solitons in trapped attractive cigar-shaped BECs has been possible under 
strong transverse binding which, in the case of weak or no axial binding,  
simulates the ideal one-dimensional situation for the formation of bright
solitons. 
The 
dimensionless nonlinear Schr\"odinger 
(NLS) equation in the attractive
or
self-focusing case  \cite{1}
\begin{equation}\label{nls}
i u_t+u_{xx}+  |u|^2u=0.
\end{equation}
sustains the following bright
 soliton \cite{1}:
\begin{eqnarray}\label{DS}
u(x,t)&=& \sqrt{2 B}\hskip 3pt \mbox{sech}
[\sqrt{B}(x-\delta+2v t)] \nonumber \\ &\times& 
\exp[-iv(x-\delta) +i(B -v v)t+i\sigma], 
\end{eqnarray}
with four parameters. The parameter  $B$ represents the amplitude as well
as pulse width, $v$ represents velocity, the parameters $\delta$ and 
$\sigma$ are phase constants. The bright soliton 
profile  
is easily recognized for $v=\delta=0$ as  
$|u(x,t)|=\sqrt{2B}\hskip 3pt  \mbox{sech} [x\sqrt{B}]$.  
There have been  experimental \cite{exdks} and theoretical \cite{thdks}
studies of the formation of  bright  solitons in a
BEC. In view of this, here we study for the first time  
the possibility of the
formation of a stable fermionic bright  soliton in a  boson-fermion
mixture.

In recent times there have been routine experimental studies on the
formation of BEC in the presence of a periodic axial optical-lattice
potential \cite{oplat}. These leads to a different condition of
trapping from the harmonic trap and generates a BEC of distinct
modulation. Hence we
also consider in this paper the modulations of the fermionic bright
solitons in the presence of an optical-lattice potential.

In Sec. II we present  the time-dependent mean-field model
consisting of a set of coupled partial differential equations involving
the bosonic and fermionic probability densities. 
 In the case
of a cigar-shaped geometry with stronger  radial trapping, 
the above model is reduced to an effective
one-dimensional form appropriate for the study of bright solitons. 
In Sec.  
III we present our results for stationary axially-free fermionic bright
solitons as well as those formed on an axial  periodic optical-lattice
potential.
We also demonstrate the stability of the bright solitons after a
perturbation  is applied. The bright solitons are found to execute
stable breathing oscillation upon perturbation. 
Finally,
a summary of our findings  is given in Sec. IV.
 
\section{Nonlinear mean-field-hydrodynamic model}

The time-dependent Bose-Einstein condensate wave
function $\Psi({\bf r},t)$ at position ${\bf r}$ and time $t $
may
be described by the following  mean-field nonlinear Gross-Pitaevskii 
equation
\cite{11}
\begin{eqnarray}\label{a} \biggr[- i\hbar\frac{\partial
}{\partial t}
-\frac{\hbar^2\nabla_{\bf r}^2   }{2m_{{B}}}
+ V_{{B}}({\bf r})
+ g_{{BB}} n_B
 \biggr]\Psi_{{B}}({\bf r},t)=0, 
\end{eqnarray}
with normalization $ \int d{\bf r} |\Psi_B({\bf r},t)|^2 = N_B. $ 
Here $m_{{B}}$
is
the mass and  $N_{{B}}$ the number of bosonic atoms in the
condensate, $n_B\equiv  |\Psi_{{B}}({\bf r},t)|^2$ is the boson 
probability density,
 $g_{{BB}}=4\pi \hbar^2 a_{{BB}}/m_{{B}} $ the strength of
inter-atomic interaction, with
$a_{{BB}}$ the boson-boson scattering length. 
The trap potential with axial symmetry may be written as  $
V_{{B}}({\bf
r}) =\frac{1}{2}m_B \omega ^2 (\rho^2+\nu^2 z^2)$ where
 $\omega$ and $\nu \omega$ are the angular frequencies in the radial
($\rho$) and axial ($z$) directions with $\nu$ the anisotropy parameter.
The probability density of an isolated 
DFG  in the Thomas-Fermi approximation  is given by
\cite{zzz}
\begin{eqnarray}\label{b}
n_F= \frac{[\mbox{max}(0,\{\epsilon_F-V_F({\bf r})\})]^{3/2}}{A^{3/2}},
\end{eqnarray}
where  $A=\hbar^2 (6
\pi^2
)^{2/3}/ (2m_F)$, $\epsilon_F$ is the Fermi energy, 
$m_F$ is the
fermionic mass, and the function $\mbox{max}$ denotes the larger of
the arguments. The confining trap potential
$V_F({\bf
r})= \frac{1}{2}m_F
\omega_F^2(\rho^2+\nu^2 z^2)$ has axial symmetry as the
bosonic potential $V_B({\bf r}),$ where $\omega_F$ is the radial
frequency. The
anisotropy parameter $\nu$ will be taken to be zero for axially-free
solitons in the following. The
number of fermionic atoms $N_F$
is given by the normalization $\int d{\bf r} n_F({\bf r})=N_F$.

We developed a set of practical time-dependent mean-field-hydrodynamic
equations for the interacting boson-fermion mixture starting from the
following Lagrangian density \cite{ska} \begin{eqnarray}\label{yy} {\cal
L}&=& \frac{i}{2}\hbar \left[ \Psi_B\frac{\partial \Psi_B^*}{\partial t} -
\Psi_B^* \frac{\partial \Psi_B}{\partial t} \right] \nonumber \\ &+&
\frac{i}{2}\hbar \left[ \sqrt{n_F}\frac{\partial {\sqrt n_F} ^*}{\partial
t} - {\sqrt n_F}^* \frac{\partial \sqrt{n_F}}{\partial t} \right]
\nonumber \\ &+& \left(\frac{\hbar^2|\nabla_{\bf r} \Psi_B|^2
}{2m_B}+V_B|\Psi_B|^2+\frac{1}{2}g_{BB} |\Psi_B|^4\right)\nonumber \\ &+&
\left(\frac{\hbar^2 |\nabla_{\bf r} \sqrt{n_F}|^2 }{6m_F}+
V_F|n_F|+\frac{3}{5} A |n_F|^{5/3}\right)\nonumber \\ &+& g_{BF} n_F
|\Psi_B|^2, \end{eqnarray} where $g_{BF}=2\pi \hbar^2 a_{BF}/m_R$ with the
boson-fermion reduced mass $m_R=m_Bm_F/(m_B+m_F),$ where $ a_{BF}$ is the
boson-fermion scattering length.

The terms in the first round bracket   on the right-hand side of
Eq. (\ref{yy}) are the standard 
Gross-Pitaevskii terms  for the bosons and are related to a
Schr\"odinger-like equation \cite{11}. However, 
terms in the second  round bracket
are derived from the hydrodynamic equation of motion of the 
fermions
\cite{capu}. Hence,
the second kinetic energy term 
has a different mass factor $6m_F$ and not the
conventional  Schr\"odinger mass factor $2m_B$ as in the first integral.
Finally, the 
last term in this equation 
 corresponds to an interaction between bosons
and fermions. 
The interaction between fermions in
spin polarized state is highly suppressed due 
to Pauli blocking 
and has been neglected in Eq. (\ref{yy})  and will be
neglected throughout this paper.

Recently, Jezek {\it et al.} \cite{jz} used the Thomas-Fermi-Weizs\"acker
kinetic energy term $T_F$ of fermions in their formulation which, in our
notation, will correspond to a fermionic kinetic energy of
$\hbar^2|\nabla_{\bf r}\sqrt{n_F}|^2/(9m_F) $ in Eq. (\ref{yy}) in place
of the present term $\hbar^2|\nabla_{\bf r}\sqrt{n_F}|^2/(6m_F) $.  This
kinetic energy term contributes little to this problem compared to the
dominating $3A|n_F|^{5/3}/5$ term in Eq. (\ref{yy}) and is usually
neglected in the Thomas-Fermi approximation.  However, its inclusion leads
to an analytic solution for the probability density everywhere
\cite{jz}. For a
discussion of these two fermionic kinetic energy terms we refer the reader
to Refs. \cite{capu,jz,pi}.

With the Lagrangian density (\ref{yy}), the Euler-Lagrange equations of
motion become \cite{ska}:  \begin{eqnarray}\label{e} \biggr[ &-&
i\hbar\frac{\partial }{\partial t} -\frac{\hbar^2\nabla_{\bf
r}^2}{2m_{{B}}} + V_{{B}}({\bf r}) + g_{{BB}}n_B \nonumber \\ &+& g_{{BF}}
n_F
 \biggr]\Psi_{{B}}({\bf r},t)=0,
\end{eqnarray}
\begin{eqnarray}\label{f} \biggr[& -& i\hbar\frac{\partial
}{\partial t}
-\frac{\hbar^2\nabla_{\bf r}^2}{6m_{{F}}}
+ V_{{F}}({\bf r})
+ A |n_F|^{2/3} \nonumber \\
&+& g_{{BF}} n_B
 \biggr]\sqrt{n_{{F}}({\bf r},t)}=0.
\end{eqnarray}

When the nonlinearity in Eq.  (\ref{f})
is  very large,
the kinetic energy term in this equation can be neglected and the
time-independent stationary form of this equation becomes
\begin{equation}\label{mod}
n_F= \frac{[\mbox{max}(0,\{\epsilon_F-V_F({\bf
r})-g_{BF}n_B\})]^{3/2}}{A^{3/2}},
\end{equation}
which is the generalization of Eq. (\ref{b}) in the presence of
boson-fermion coupling. Equation (\ref{mod}) has been used by 
Modugno
{\it et al.} \cite{zzz} for an analysis of a DFG-BEC mixture. 
In actual experimental
condition the nonlinearity in Eq.  (\ref{f})    is quite large and  Eq.
(\ref{mod}) is a good approximation.

For the study of bright  solitons
we shall reduce Eqs. (\ref{e}) and  (\ref{f}) to the minimal 
 one-dimensional form under the action of stronger radial trapping. 
The one-dimensional form is appropriate for     studying bright 
solitons in the so-called cigar-shaped quasi-one-dimensional geometry
where 
$\nu << 1$. For radially-bound and axially-free solitons we eventually set 
$\nu =0$.
In this case the dynamical equations can be reduced to strict
one-dimensional coupled NLS equations  without any trap. 
We perform this
reduction below where we 
take $V_B({\bf r})=V_F({\bf r})= \frac{1}{2}m_B\omega^2(
\rho^2+\nu^2 z^2)$ which corresponds to a
reduction of $\omega_F$ and
$\nu  \omega_F$ in $V_F({\bf r})$
by a factor
$\sqrt{m_B/m_F}$ as in the study by 
Modugno {\it et al.} \cite{zzz} and Jezek  {\it et al.} \cite{jz}.

For  $\nu =0$, Eqs. (\ref{e})
and (\ref{f}) can be reduced to an effective 
one-dimensional form by considering 
solutions of the type 
$\Psi_B({\bf r},t)=  \phi_B(z,t)\psi_B^{(0)}( \rho)$ and
$\sqrt{n_F({\bf r},t)}=  \phi_F(z,t)\psi_F^{(0)}( \rho),$
where
\begin{eqnarray}\label{wfx}
|\psi_i^{(0)}(\rho)|^2&\equiv&
{\frac{M_i\omega}{\pi\hbar}}\exp\left(-\frac{M_i
\omega
\rho^2}{\hbar}\right),
\end{eqnarray}
and  $i=B,F$ represents bosons and fermions and $M_B=m_B$ and
$M_F=\sqrt{3m_Bm_F}. $ The expression (\ref{wfx})
corresponds to the respective
ground state wave function in the absence of nonlinear interactions and
satisfies
\begin{eqnarray}
-\frac{\hbar^2}{2m_B}\nabla_\rho ^2\psi_B^{(0)}
+
\frac{1}{2}m_B\omega^2\rho^2
\psi_B^{(0)}&=&\hbar\omega
\psi_B^{(0)},\\
-\frac{\hbar^2}{6m_F}\nabla_\rho^2\psi_F^{(0)}+
\frac{1}{2}m_B\omega^2\rho
^2\psi_F^{(0)}&=&\sqrt{\frac{m_B}{3m_F}}\hbar\omega
\psi_F^{(0)},\nonumber \\
\end{eqnarray}
with normalization 
$2\pi \int_{0}^\infty |\psi_i^{(0)}(\rho)|^2 \rho d\rho=1.$
Now the dynamics is carried by $ \phi_i(z,t)$ and the radial dependence is
frozen in the ground state $\psi_i^{(0)}(\rho)$.
The factorization of $\Psi_B$ and $\sqrt{n_F}$ above follows from the
structure of the mathematical equations (\ref{e}) and (\ref{f}).
Although $n_F$ gives the probability density of DFG it may not be to the
point to
associate $\psi_F^{0}$ and $\phi_F$ to physical fermionic one-particle
wave functions. The true fermionic wave function has the form of a
many-particle 
Slater determinant. Nevertheless,  
the functions $\psi_F^{0}$ and $\phi_F$ 
could be regarded as mathematical functions related to
fermionic density \cite{yyy1}. In the quasi-one-dimensional cigar-shaped
geometry the
linear fermionic and bosonic probability densities are given by
$|\phi_F(z,t)|^2$ and
$|\phi_B(z,t)|^2,$ respectively.

Averaging over the radial mode $\psi_i^{(0)}(\rho)$, 
i.e., multiplying
Eqs. (\ref{e}) and (\ref{f})
by  $\psi_i^{(0)*}(\rho)$
and integrating over $\rho$, we obtain the following one-dimensional 
dynamical equations \cite{abdul}:
\begin{eqnarray}\label{i} \biggr[ &-& i\hbar\frac{\partial
}{\partial t}
-\frac{\hbar^2}{2m_{{B}}}\frac{\partial^2}{\partial z^2}
 \nonumber \\  &+&F_{BB}|
\phi_B|^2
+ F_{BF}| \phi_F|^2
 \biggr] \phi_{{B}}(z,t)=0, 
\end{eqnarray}
\begin{eqnarray}\label{j} 
\biggr[& -& i\hbar\frac{\partial
}{\partial t}
-\frac{\hbar^2}{6m_F}\frac{\partial^2}{\partial z^2}
\nonumber \\  
&+&F_{FF}|
\phi_F|^{4/3}  
+ F_{BF}| \phi_B|^2
 \biggr] \phi_{{F}}(z,t)=0, 
\end{eqnarray}
where 
\begin{eqnarray}
 F_{BB}=g_{BB}\frac{\int_0^\infty|\psi_B^{(0)}|^4\rho d\rho}
{\int_0^\infty|\psi_B^{(0)}|^2\rho d\rho}=
g_{BB}{\frac{m_B\omega}{2\pi\hbar}},
\end{eqnarray}
\begin{eqnarray}
F_{BF}=g_{BF}\frac{\int_0^\infty|\psi_F^{(0)}|^2|\psi_B^{(0)}|^2\rho 
d\rho}{\int_0^\infty|\psi_B^{(0)}|^2\rho d\rho}
=g_{BF}{\frac{M_{BF}\omega}{\pi\hbar}},
\end{eqnarray}
\begin{eqnarray}
F_{FF}=A
\frac{\int_0^\infty|\psi_F^{(0)}|^{2+4/3}\rho
d\rho}{\int_0^\infty|\psi_B^{(0)}|^2\rho
d\rho} =
{\frac{3A}{5}}\left[
\frac{M_F\omega}{\pi \hbar}    \right]^{2/3}.
\end{eqnarray}
In Eqs. (\ref{i}) and (\ref{j}) 
the normalization 
is given by $\int_{-\infty}^\infty |\phi_i(z,t)|^2
dz = N_i$. In these equations we have set the anisotropy parameter 
$\nu=0$
 to remove the axial trap and thus to generate axially-free
quasi-one-dimensional solitons.

For calculational purpose it is convenient to reduce 
the set  (\ref{i}) and (\ref{j})  to
dimensionless form 
by introducing convenient  dimensionless variables. Although the algebra is
quite straightforward, the expressions become messy with  different
factors
of masses. 
As we shall not be interested in a particular
boson-fermion 
system in this
paper, but will be concerned with the formation of fermionic bright 
solitons 
in general,  we take in the rest of this paper $m_B=3 m_F=
m({^{87}\mbox{Rb}})$, where $m({^{87}\mbox{Rb}})$ is the mass of
$^{87}$Rb, and 
whence $m_R=3m_F/4, M_B=M_F=m_B,$ and
$M_{BF}=m_B/2. $
In
the two experimental
situations of Refs. \cite{exp4,exp5} 
$m_B \approx 
3m_F$.

In Eqs. (\ref{i}) and (\ref{j}),   
we consider the dimensionless variables 
$\tau=t \omega/2$,
$y=z /l$,
${\chi}_i=
\sqrt{(l/N_i)} \phi_i$, with $l=\sqrt{\hbar/( \omega m_B)}$, 
so that 
\begin{eqnarray}\label{m} \biggr[& - &i\frac{\partial
}{\partial \tau}
-\frac{d^2}{dy^2} 
+    
 N_{BB}
\left|{{\chi}_B}\right|^2\nonumber \\        
&+&N_{BF}
  \left|{{\chi}_F}\right|^2                  
 \biggr]{\chi}_{{B}}({y},\tau)=0,         
\end{eqnarray}
\begin{eqnarray}\label{n} \biggr[& - & i\frac{\partial
}{\partial \tau}-\frac{d^2}{dy^2} 
+
N_{FB}
  \left|{{\chi}_B} \right|^2
  \nonumber \\
&+&
N_{FF}
  \left|{{\chi}_F}
\right|^{4/3}
 \biggr]{\chi}_{{F}}(y,\tau)=0,
\end{eqnarray}
where
$N_{BB}=4a_{BB}N_B/l,$
$N_{BF}=8a_{BF}N_F/l,$ 
$N_{FB}=8a_{BF}N_B/l,$ and
$N_{FF}=9(6\pi N_F)^{2/3}/5. $
 In Eqs.
(\ref{m}) and (\ref{n}),
the normalization condition  is given by 
$\int_{-\infty}^\infty |\chi_i(y,\tau)|^2 dy =1 .$
Equations (\ref{m}) and (\ref{n}) are the coupled one-dimensional 
NLS equations describing the formation of solitons in the DFG-BEC mixture 
in a cigar-shaped quasi-one-dimensional geometry.

In Eqs. (\ref{m}) and (\ref{n}) the term $N_{FF}|\chi_F|^{4/3}$ represents
a very strong Pauli repulsion which increases with the fermion number
$N_F$. The purpose of this study is to show that a sufficiently strong  
attractive boson-fermion coupling
term $N_{FB}|\chi_B|^2$ can  overcome
this repulsion and  form the bright solitons.

\section{Numerical Result}

We solve the coupled mean-field-hydrodynamic  equations 
 (\ref{m}) and
(\ref{n}) for bright  solitons
numerically using a time-iteration
method based on the Crank-Nicholson discretization scheme
elaborated in Ref. \cite{sk1}.  
We
discretize the mean-field-hydrodynamic  equation
using time step $0.0005$ and space step $0.025$.

We performed  the time evolution of the set of equations (\ref{m}) and 
(\ref{n}) introducing harmonic oscillator potentials $y^2$ in these
equations and setting the nonlinear terms to
zero: $N_{BB}=N_{BF}=N_{FB}=N_{FF}=0$ and  
starting with the eigenfunctions of the linear harmonic
oscillator problem, e.g., with $\chi_B(y,\tau)=\chi_F(y,\tau)=
\pi^{-1/4}\exp(-y^2/2)\exp(-i\tau).$  The introduction of the extra
harmonic oscillator potential in these equations 
only aids in starting the time evolution
with an exact analytic
solution. In the end the harmonic
oscillator potentials will be set equal to zero  and will have no effect
on
the final wave function for solitons. 
During the course of time evolution the nonlinear
terms are  switched on very slowly and the resultant solutions iterated 
until convergence was obtained. Then the time evolution is continued
and the harmonic oscillator potential
terms in both bosonic and fermionic equations are  slowly switched off by
reducing the $y^2$ term to zero in 10000 steps of time evolution.
Then 
the resultant solutions are iterated about 50000 times 
for
convergence without any harmonic oscillator potential. 
If converged
solutions are obtained, they correspond to the
required axially-free bright solutions in the absence of any axial
potential.   
In our numerical investigation 
as in the theoretical study of Refs. \cite{jz,ska} 
we use  $\omega = 2\pi
\times
100$ Hz, and
take $m_B$ as the mass of $^{87}$Rb. Consequently, the unit of
length $l\approx1$ $\mu$m and unit of time $2/\omega \approx 3$ ms.

\begin{figure}
 
\begin{center}
\includegraphics[width=1.\linewidth]{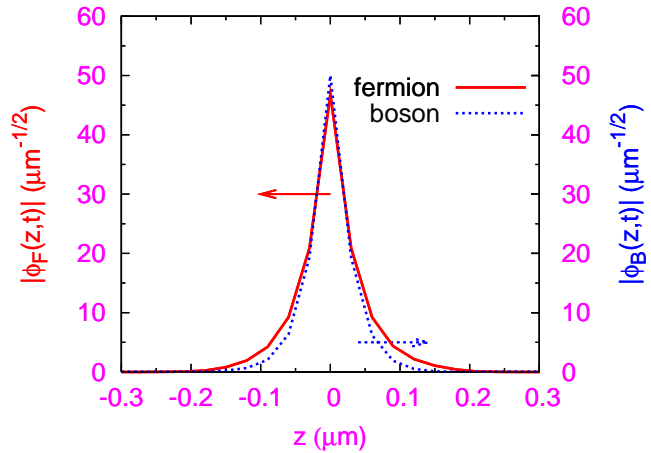}
\end{center}

\caption{(Color online)  The stationary 
 function $|\phi_i(z,t)|$
 for axially-free bosonic (dotted line) and
fermionic (full line) bright solitons  vs. $z$ for 
 $N_F=N_B=1000$, $a_{BB}=5 $ nm,  $a_{BF}=-20 $ nm,  harmonic
oscillator length $l\approx 1$ $\mu$m and $\nu =0$.  
The arrows in dotted
and full lines indicate the bosonic and fermionic axes, respectively. 
The nonlinearity
parameters are
$N_{BB}=20 $, $N_{BF}= -160$, $N_{FB}= -160$, and $N_{FF}=274.6$.
}
\end{figure}

\begin{figure}
 
\begin{center}
\includegraphics[width=1.\linewidth]{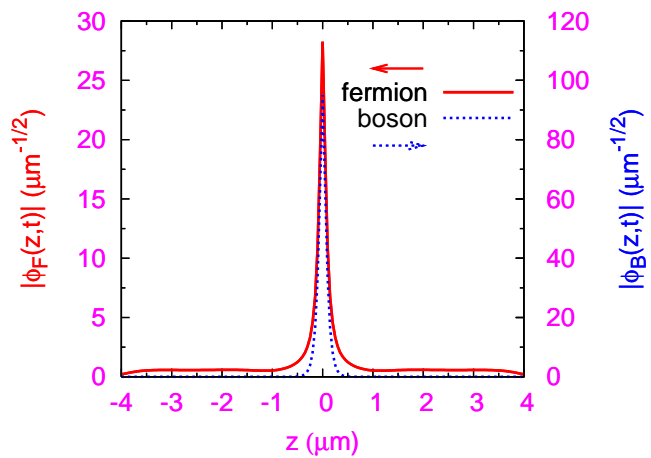}
\end{center}

\caption{(Color online)  The stationary 
function $|\phi_i(z,t)|$
 for axially-free bosonic (dotted line) and
fermionic (solid line) bright solitons  vs. $z$ for 
 $N_F=1000$, $N_B=10000$, $a_{BB}=-1 $ nm,  $a_{BF}=-1.875 $ nm, harmonic
oscillator length $l\approx 1$ $\mu$m  and $\nu =0$. The arrows in dotted
and full lines indicate the bosonic and fermionic axes, respectively. 
 The nonlinearity
parameters are
$N_{BB}=-40 $, $N_{BF}= -15$, $N_{FB}= -150$, and $N_{FF}=274.6$.
}
\end{figure}

First we solve Eqs. (\ref{m}) and (\ref{n}) with  
$N_F=N_B= 1000$, $a_{BB}=5$ nm  and
$a_{BF}=-20$ 
nm. 
This value of  $a_{BB}$ is the
experimental scattering length of Rb atoms \cite{11}, and  $a_{BF}=-20$
nm is the experimental 
scattering
length of the Rb-K system \cite{exp5,exp5x}. With these
parameters the nonlinearities in Eqs. (\ref{m}) and (\ref{n}) are
$N_{BB}=20 $, $N_{BF}= -160$, $N_{FB}= -160$, and $N_{FF}=274.6$.
The converged  bright solitons are plotted in Fig. 1. In this
case the
fermionic and bosonic functions, $\phi_F$ and $\phi_B$, respectively,   
have similar spatial extentions.
It is possible to have solitons with different 
extensions in space by varying the parameters of the system. 
We took the experimental values for the scattering lengths in
Fig. 1. However,  the scattering length can be manipulated in the
boson-fermion
$^6$Li-$^{23}$Na and $^{40}$K-$^{87}$Rb systems
near the recently
discovered Feshbach resonances in them \cite{fesh} by varying a background
magnetic field. Thus by varying the scattering length and the number of
atoms we could arrive at different values of nonlinearity than in
Fig. 1.

To simulate a different situation of nonlinearity parameters next we take
$a_{BB}=-1$ nm, $a_{BF}=-1.875$ nm, $N_F=1000$, and $N_B=10000$. so that
$N_{BB}=-40 $, $N_{BF}= -15$, $N_{FB}= -150$, and $N_{FF}=274.6$. In this
case the profiles of the bright solitons shown in Fig. 2 are very
different from those in Fig. 1. In Fig. 1 both the solitons are localized
to a small region in space, whereas in Fig. 2 only the bosonic soliton is
localized to a small region in space whereas the fermionic soliton extends
to a very large region of space.

\begin{figure}
 
\begin{center}
\includegraphics[width=1.\linewidth]{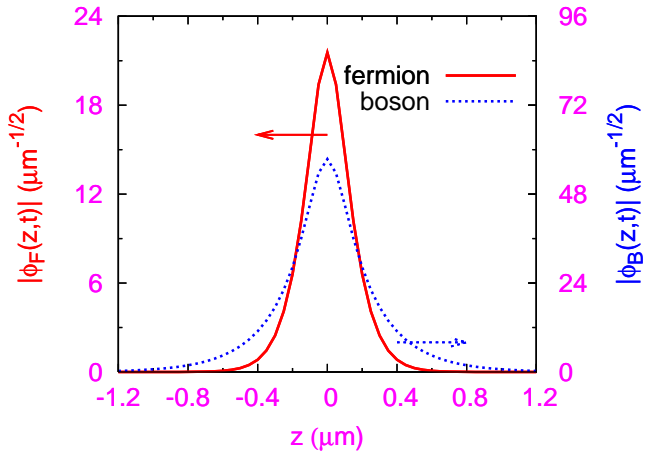}
\end{center}

\caption{(Color online)  
The stationary 
 function $|\phi_i(z,t)|$
 for axially-free bosonic (dotted line) and
fermionic (solid line) bright solitons  vs. $z$ for 
 $N_F=1000$, $N_B=10000$, $a_{BB}=0.5 $ nm,  $a_{BF}=-3.75 $ nm,  $\nu
=0$
and
harmonic
oscillator length $l\approx 1$ $\mu$m.  The arrows in dotted
and full lines indicate the bosonic and fermionic axes, respectively. 
The nonlinearity parameters are
$N_{BB}=20 $, $N_{BF}= -30$, $N_{FB}= -300$, and $N_{FF}=274.6$.
}
\end{figure}

\begin{figure}
 
\begin{center}
\includegraphics[width=\linewidth]{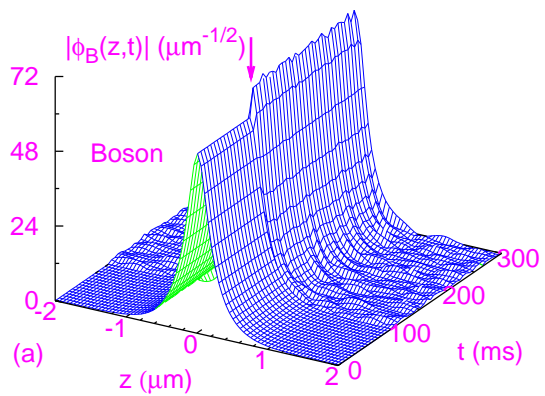}
\includegraphics[width=\linewidth]{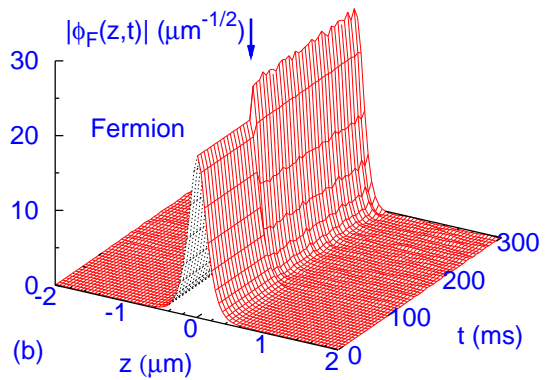}
\end{center}

\caption{(Color online)  The 
 function $|\phi_i(z,t)|$
 for (a) bosonic and
(b) fermionic bright solitons  vs. $z$  and $t$ for
the solitons of Fig. 3.
At $t=0$ 
 $N_F=1000$, $N_B=10000$, $a_{BB}=0.5 $ nm,  $a_{BF}=-3.75 $ nm,  $\nu
=0$
and
harmonic
oscillator length $l\approx 1$ $\mu$m.  The nonlinearity parameters at
$t=0$ are
$N_{BB}=20 $, $N_{BF}= -30$, $N_{FB}= -300$, and $N_{FF}=274.6$. At
$t=100$ ms (marked by arrows) the bright solitons are set into small
breathing oscillations by suddenly jumping the nonlinearities $N_{BF}$ and
$N_{FB}$ to $-33$ and $-330$, respectively. } \end{figure}

Next we consider  $N_F=1000$, $N_B=10000$, $a_{BB}=0.5 $ nm
and
$a_{BF}=-3.75 $ nm, so that the nonlinearity parameters are 
$N_{BB}=20 $, $N_{BF}= -30$, $N_{FB}= -300$, and $N_{FF}=274.6$.
The profiles of the solitons in Fig. 3 are different from those in Figs. 1
and 2. In this case the bosonic  function extends over  a longer 
region in space than the fermionic  function. In Figs. 1 and 2 it was
the fermionic  function that extends over a longer region in space.

In Fig. 2 the nonlinearity $N_{FB}$
appearing in the fermion equation is less attractive compared to that in
Figs. 1 and 3. Hence the resultant  nonlinear interaction in the fermion
component 
is less
attractive and hence the fermionic soliton extends to a large distance in
space. If the attraction in  $N_{FB}$   is further reduced the fermionic
soliton ceases to bind.

Hence by manipulating the parameters one could have different situations 
of localization of the solitons. One could either have both the solitons
extending up to similar distance in space as in Fig. 1 or one of the
solitons extending over a longer region in space as in Figs. 2 and 3. It
is worth emphasizing that in the fermionic equation (\ref{n}) the diagonal
nonlinearity $N_{FF}$ is repulsive, hence the binding solely comes from
the attractive off-diagonal nonlinearity $N_{FB}$  corresponding to an
attractive boson-fermion interaction. Hence for a fermionic soliton to
appear  $N_{FB}$ and $a_{BF}$ are always taken to be negative or
attractive. Consequently,   $N_{BF}$ is also negative. Finally, terms
$N_{BB}$ and $a_{BB}$ could be  either  positive or negative. When these
terms are positive or repulsive the bosonic solitons are formed due to an
attractive or negative $N_{BF}$, as in Fig. 3.

Next we study the stability of the bright solitons numerically. We
consider the soliton of Fig. 3 and during time evolution we suddenly jump
at $t=100$ ms the nonlinearity $N_{BF}$ from $-30$ to $-33$ and the
nonlinearity $N_{FB}$ from $-300$ to $-330$. This can be achieved by
manipulating a
background magnetic field near 
a
Feshbach resonance \cite{fesh} in the boson-fermion interaction and
thus varying the boson-fermion scattering length by 10$\%$. Due to
the sudden change in the nonlinearity the bosonic and  fermionic bright
solitons are set into stable breathing oscillation.  The evolution of the
wave function profile of the two solitons are shown in Figs. 4. 
The solitons are found to execute stable non-periodic breathing
oscillation. In Fig. 5 
we plot the root-mean-square size $\langle z \rangle_{rms}$ of the bosonic
and fermionic
solitons  of Fig. 4 as a function of time. The breathing
oscillation of the two solitons after the perturbation is applied results
in the stable non-periodic oscillation of the root-mean-square sizes
illustrated in Fig. 5. As, after applying the perturbation, the 
boson-fermion attraction has been increased this corresponds to a 
reduction in the root-mean-square size $\langle z \rangle_{rms}$ as we 
find in Fig. 5.  The steady propagation of the solitons in Figs. 4 and the 
stable oscillation of their root-mean-square sizes in Fig. 5
after the
perturbation is
applied demonstrate the stability of the solitons.

\begin{figure}
 
\begin{center}
\includegraphics[width=\linewidth]{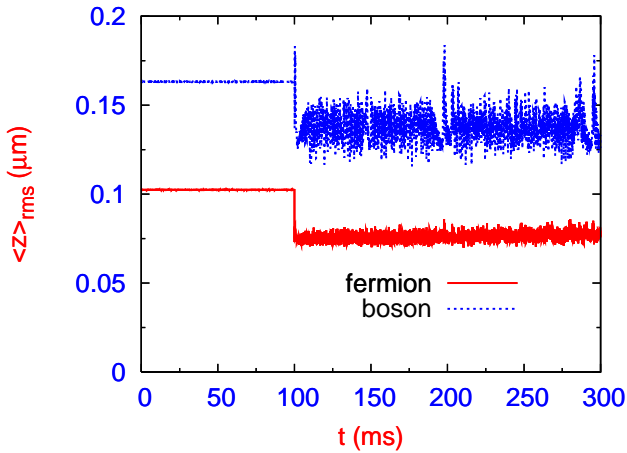}
\end{center}

\caption{(Color online)  The root-mean-square size $\langle z
\rangle_{rms}$ of the bosonic (dotted line)
and fermionic (full line)
solitons  of Fig. 4 vs. time. } \end{figure}

Finally, we consider the fermionic bright solitons formed on a periodic
optical-lattice potential. For that purpose we include in Eqs. (\ref{m})
and (\ref{n}) the following optical-lattice potential formed by a
standing-wave laser beam \cite{oplat} 
\begin{equation}
V_{\mbox{\small{OP}}} = V_0 \sin^2(2\pi  y/\lambda),
\end{equation}
where $V_0$ is the strength, and $\lambda$ is the wave length of the
laser. In our calculation we take $V_0=100$ and $\lambda = \pi/2$. To
solve Eqs. (\ref{m}) and (\ref{n}) with this optical-lattice potential and
desired nonlinearities, we again start the time evolution with the
solution of the linear oscillator problem. In the course of time evolution
the nonlinearities and the optical-lattice potential are slowly
introduced and eventually the harmonic oscillator potential is slowly
removed.  Then the final solutions are iterated for convergence. 
 The
resultant soliton wave functions are plotted in Fig. 6 for $N_F= 1000$,
$N_B=10000$, $a_{BB}=0.3$ nm, $a_{BF}=-2.375$ nm.  
The optical-lattice potential introduces modulations in the solitonic 
wave function. For
the parameters of Fig. 6 the modulations are more prominent on the
fermionic soliton than the bosonic one. By changing the
parameters it is
possible to have modulations on the bosonic soliton as well. 

\begin{figure}[!ht]
 
\begin{center}
\includegraphics[width=\linewidth]{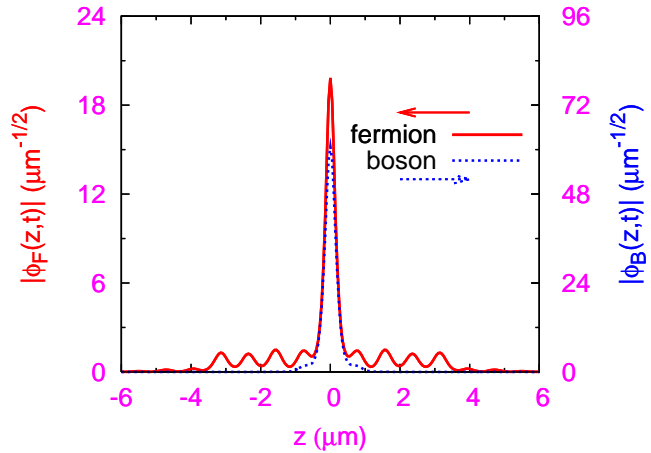}
\end{center}

\caption{(Color online)  
The stationary 
 function $|\phi_i(z,t)|$
 for bosonic (dotted line) and
fermionic (full line) bright solitons  vs. $z$ for 
 $N_F=1000$, $N_B=10000$, $a_{BB}=0.3 $ nm,  $a_{BF}=-2.375 $ nm,  $\nu
=0$
and
harmonic
oscillator length $l\approx 1$ $\mu$m in the presence of the
optical-lattice potential $V(y)=100 \sin^2 (4 y)$. The arrows in dotted
and full lines indicate the bosonic and fermionic axes, respectively. 
 The nonlinearity
parameters
are $N_{BB}=12 $, $N_{BF}= -19$, $N_{FB}= -190$, and $N_{FF}=274.6$.
} \end{figure}

\section{Summary}
 
We use a coupled set of time-dependent mean-field-hydrodynamic equations
for a boson-fermion mixture to study the formation of fermionic bright
soliton in a DFG as a stationary state. In this study we take the
boson-boson interaction to be both attractive and repulsive and the
boson-fermion interaction to be attractive. An attractive boson-fermion
interaction is necessary for the formation of a fermionic bright soliton
as the diagonal nonlinearity $N_{FF}$ in the fermion-fermion system is
always repulsive.

In the present study we demonstrate that stable solitons can be formed 
in coupled NLS equations for the boson-fermion mixture
with repulsive diagonal nonlinearities  and
attractive off-diagonal nonlinearities above some cut-off values. In
another study \cite{pla} we showed the possibility of the formation of
bright 
solitons in coupled bosonic condensates with intraspecies repulsion
supported by interspecies
attraction. 
The stability of the present fermionic and bosonic solitons is
demonstrated through
their sustained breathing oscillation  initiated by a sudden jump
in the boson-fermion scattering length. 
Bright solitons have been created experimentally
in attractive
BECs in three dimensions in the presence of  radial trapping only without
any
axial trapping \cite{exdks}. 
In view of this fermionic bright solitons can be observed in the
laboratory in the presence of  radial trapping only in a mixture of a DFG 
and BEC. We also suggest the possibility of the formation of fermionic  
solitons
on a periodic optical-lattice potential.  
In the present investigation  we used a set of mean-field equations for 
the DFG-BEC mixture. A proper treatment of the DFG should be performed 
using a fully antisymmetrized many-body Slater determinant wave
function. However, we do not believe that the present conclusion about the
existence of robust fermionic solitons in a DFG-BEC mixture to be so
peculiar as to have no general validity.

\acknowledgments

The work is supported in part by the CNPq 
of Brazil.


\end{document}